\documentclass{mn2e}
\usepackage{graphicx}

\def\go{
\mathrel{\raise.3ex\hbox{$>$}\mkern-14mu\lower0.6ex\hbox{$\sim$}}
}
\def\lo{
\mathrel{\raise.3ex\hbox{$<$}\mkern-14mu\lower0.6ex\hbox{$\sim$}}
}
\def\simeq{
\mathrel{\raise.3ex\hbox{$\sim$}\mkern-14mu\lower0.4ex\hbox{$-$}}
}

\def\etal{{et al.\ }}

\def\msun{{\rm M_{\odot}}}

\begin{document}

\title[The X-ray and UV evolution of V2491 Cyg]
{{\it Swift} observations of the X-ray and UV evolution of V2491 Cyg (Nova Cyg 2008 No. 2)}
\author[K.L. Page \etal]{K.L. Page$^{1}$, J.P. Osborne$^{1}$, P.A. Evans$^{1}$, G.A. Wynn$^{2}$, A.P. Beardmore$^{1}$, \and R.L.C. Starling$^{1}$, M.F. Bode$^{3}$, A. Ibarra$^{4}$, E. Kuulkers$^{5}$, J.-U. Ness$^{4}$ \& G.J. Schwarz$^{6}$ \\
$^{1}$ X-Ray and Observational Astronomy Group, Department of Physics \&
  Astronomy, University of Leicester, LE1 7RH, UK\\
$^{2}$ Theoretical Astrophysics Group, Department of Physics \& Astronomy, University of Leicester, LE1 7RH, UK\\
$^{3}$  Astrophysics Research Institute, Liverpool John Moores University, CH41 1LD, UK\\
$^{4}$ XMM-Newton Science Operations Centre, ESAC, Apartado 78, 28691 Villanueva de la Ca{\~n}ada, Madrid, Spain
\\
$^{5}$ INTEGRAL Science Operations Centre, ESAC, Apartado 78, 28691 Villanueva de la Ca{\~n}ada, Madrid, Spain
\\
$^{6}$ West Chester University, Department of Geology and Astronomy, West Chester, PA 19383, USA\\
}

\label{firstpage}

\maketitle

\begin{abstract}

We present extensive, high-density {\it Swift} observations of V2491~Cyg (Nova Cyg 2008 No. 2). Observing the X-ray emission from only one day after the nova discovery, the source is followed through the initial brightening, the Super-Soft Source phase and back to the pre-outburst flux level. The evolution of the spectrum throughout the outburst is demonstrated.
The UV and X-ray light-curves follow very different paths, although changes occur in them around the same times, indicating a link between the bands. Flickering in the late-time X-ray data indicates the resumption of accretion.

We show that if the white dwarf is magnetic, it would be among the most magnetic known; the lack of a periodic signal in our later data argues against a magnetic white dwarf, however. We also discuss the possibility that V2491~Cyg is a recurrent nova, providing recurrence timescale estimates.

\end{abstract}

\begin{keywords}
stars: individual: V2491 Cyg --- novae, cataclysmic variables 
\end{keywords}

\date{Received / Accepted}

\section{Introduction}
\label{intro}

A nova occurs in an interacting binary system consisting of
a white dwarf (WD) primary and a lower mass secondary star; the actual nova explosion happens when
enough matter is transferred (via Roche lobe overflow or wind
accretion) to the surface of the WD such that the pressure and temperature at the base
of the envelope of accreted material is sufficient to trigger a
thermonuclear runaway (TNR). For a review, see Bode \& Evans (2008). Part of the envelope of material is
expelled and obscures the X-rays emitted from the WD surface. As the
ejected material expands further, it eventually becomes optically thin and the surface nuclear burning becomes visible,
with emission peaking in the soft X-ray band -- the so-called
Super-Soft Source (SSS) state (Krautter 2008). Nuclear burning is expected to continue at constant
bolometric luminosity, depleting the envelope mass and causing the envelope (and
the effective radius of the photosphere) to contract. Eventually
nuclear burning can no longer be sustained and the nova returns to
quiescence.



V2491 Cyg was detected in nova outburst on 2008 April 10.728 (Nakano 2008; Ayani
\& Matsumoto 2008) -- the first time this source has been seen to
erupt. The {\it Swift} X-ray
Telescope (XRT; Burrows et al. 2005) had previously observed this
field as part of a follow-up survey aimed at identifying X-ray
counterparts for BAT (Burst Alert Telescope; Barthelmy et
al. 2005) survey sources (Tueller et al. 2008). These earlier observations are discussed by Ibarra et al. (2009).

Observations following the nova explosion were obtained in the radio (Kuulkers et al. 2008), infrared (Ashok et al. 2008; Naik, Banerjee \& Ashok 2009; Lynch et al. 2008a; Rudy et al. 2008), optical (Tomov et al. 2008a,b; Ayani \& Matsumoto 2008; Schmeer \& Hornoch 2008; Nakano 2008, Waagen et al. 2008; Balman, Pekon \& Kiziloglu 2008; Baklanov, Pavlenko \& Berezina 2008; Helton et al. 2008; Urbancok et al. 2008), UV and X-ray (Ibarra et al. 2008; Kuulkers et al. 2008; Page et al. 2008a; Osborne et al. 2008; Ness et al. 2008b,c) bands.
Here we present the very extensive {\it Swift} observations of the X-ray and UV evolution of V2491~Cyg from one day after the nova discovery, through the super-soft phase and back to approximately quiescence. This high-quality dataset allows us to confront the nova models in excellent detail.

\section{Data Reduction and Results}
\label{res}

The {\it Swift} data were processed and analysed using version 2.8 of the software (released on 2007-12-06, as part of HEASoft 6.4)\footnote{http://swift.gsfc.nasa.gov/docs/swift/news/2007/\\swift$\_$softwarev2.8$\_$caldb.html}, following the standard methods. At all times, the XRT bad columns (caused by micrometeoroid damage in May 2005; Abbey et al. 2005) were corrected for by the use of exposure maps. When there were fewer than $\sim$~20 source counts in a single observation, Bayesian statistics were used to calculate the uncertainties on the X-ray count rates (Kraft, Burrows \& Nousek 1991) and hardness ratios (Park et al. 2006). The hardness ratio is defined here as the ratio of the counts in the 1.5--10~keV and 0.3--1.5~keV bands. 

Starting a week after the nova outburst, the UVOT (UV/Optical Telescope; Roming et al. 2005) collected data in the uvw2 filter (central wavelength of 1928 \AA, FWHM~=~657 \AA), using a combination of image and event mode; earlier observations had the UVOT filter wheel in the blocked position. 
The position of the source in each observation needs to be precisely known to obtain the correct photometry automatically. However, many of the UVOT images could not be astrometrically corrected using the standard tools, because the USNO-B1 field is extremely crowded; instead, {\sc uvotdetect} was run on each image, assuming that V2491~Cyg was the source closest to its catalogued position and within 10~arcsec of that location. (The field had been previously inspected to confirm there were no sources within 10~arcsec of V2491~Cyg.) A sample of these centroids were checked by hand, to confirm their accuracy.
A 5~arcsec radius circular region was defined, centred on the point, to ensure the photometry was correctly calibrated, and an appropriate background region selected. {\sc uvotsource} was then used to extract the light-curve information. The UVOT light-curve is shown in Figure~\ref{bands}. Before about day 20 after the outburst the source was bright enough in the UV to saturate the UVOT detector, so these measurements have not been included in the plot.

\subsection{Pre-nova source}
\label{pre}

\begin{figure*}
\begin{center}
\includegraphics[clip, angle=-90, width=18cm]{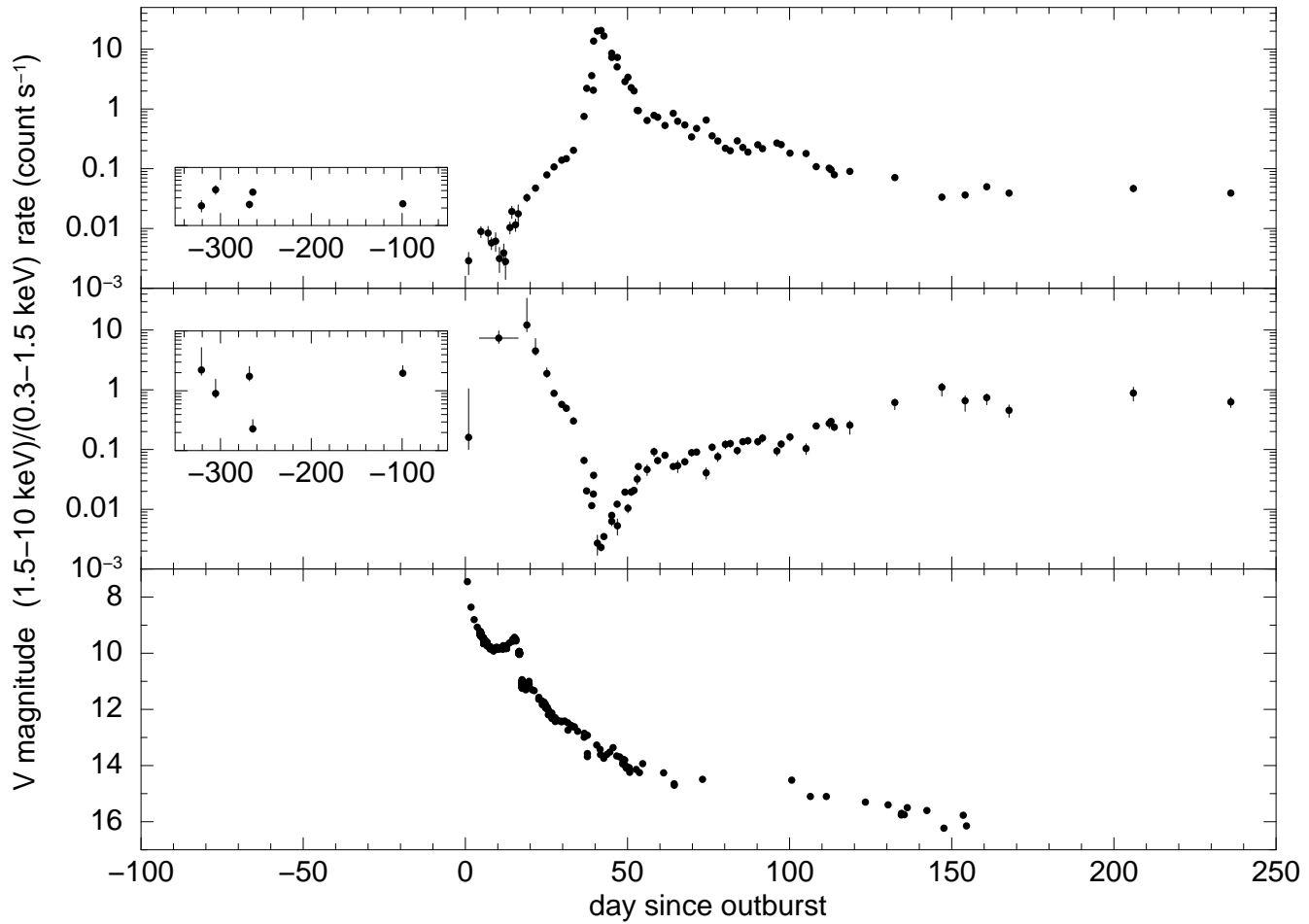}
\caption{Top two panels: {\it Swift}-XRT count-rate (0.3--10~keV) and hardness ratio before, during and after the outburst of V2491~Cyg. The insets show the count-rate (top) and hardness ratio (bottom) before outburst (Ibarra et al. 2009) on the same vertical scale as for the main panel. The bottom panel shows the $V$-band evolution from the American Association of Variable Star Observers (AAVSO).}
\label{1bin}
\end{center}
\end{figure*}

A variable X-ray source has been seen at the position of V2491~Cyg since the {\it ROSAT} survey era (1990/91), with the faint source 1RXS J194259.9+321940 having a PSPC count-rate of 0.028~$\pm$~0.011 count~s$^{-1}$ (Voges et al. 2000); this corresponds to 0.03 count s$^{-1}$ in the XRT which, as shown in the inset of Figure~\ref{1bin}, is approximately the pre-nova XRT count rate measured in the months before outburst. In addition, the {\it ROSAT}/PSPC source 2RXP~J194302.0+321912 is in agreement with the optical nova coordinates (RA: 19$^h4$3$^m$01.96$^s$, Dec = +32$^o$19'13.8''; Nakano et al. 2008). The XMM-Newton slew-survey source XMMSL1 J194301.9+321911 is also consistent with the position of V2491~Cyg, although the source was detected at less than the 4$\sigma$ level (Ibarra \& Kuulkers 2008; Ibarra et al. 2008). Archival plates show that there was a persistent (measured over $\sim$~16~yr) optical source of $V$~$\sim$~17.1 (Jurdana-Sepic \& Munari 2008) at the position of V2491~Cyg.

The inset panels of Figure~\ref{1bin} show the early (pre-nova) XRT data for this source. Both the count-rate and hardness ratio are variable, with the fourth observation being noticeably softer than the others (though, as will be shown later, still harder than the super-soft state). More details on the pre-nova are given in Ibarra et al. (2009)


\subsection{Post-nova source}
\label{post}

\subsubsection{X-ray data}
\label{post:x}

Following the discovery of V2491~Cyg (Nova Cyg 2008 No. 2; Nakano 2008; Ayani \& Matsumoto 2008), a {\it Swift} Target of Opportunity (ToO) observation was requested and the field observed less than a day later. An X-ray source was identified, though the count-rate was almost an order of magnitude fainter, and somewhat softer, than during the previous serendipitous observations (Figure~\ref{1bin}). After this initial ToO, regular {\it Swift} observations were scheduled to follow the evolution of the nova.



\begin{figure*}
\begin{center}
\includegraphics[clip, angle=0, width=16cm]{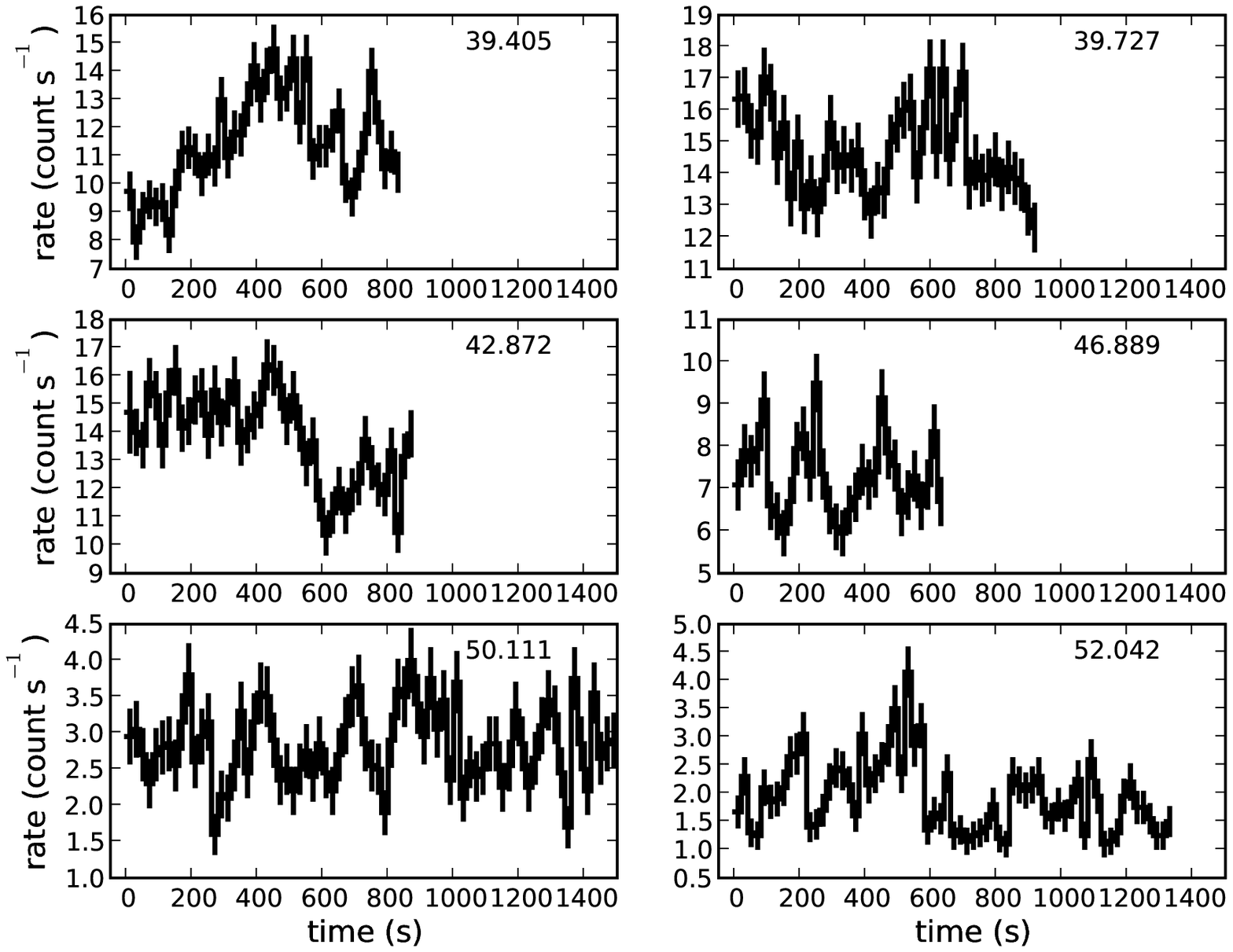}
\caption{Examples of short term variability in the 0.3--10~keV X-ray data of V2491~Cyg. The numbers given at the top right of each panel indicate the day after outburst corresponding to T=0 in that panel. The vertical scaling is different for the separate panels.}
\label{wiggles}
\end{center}
\end{figure*}

The X-ray source started to brighten, before fading again over an interval of about a week. Around 14 days after the nova explosion, the source began to brighten steadily, shortly after the start of a rebrightening in the optical (Figure~\ref{1bin}). A sudden, rapid increase in count-rate occurred between days 36 and 42 after the outburst; at this time, the nova evolved into a super-soft source. The X-ray source then started to fade until day 57, at which point the rate of decay decreased. The fading trend has continued, although there have been brief periods of slight rebrightening; this is discussed in more detail in Section~\ref{flick}. 
In some light-curve segments, short time variations are seen. Figure~\ref{wiggles} shows a couple of cycles of a $\sim$~200~s modulation, however no strict periodicity was revealed by a power-spectral analysis. After $\sim$~day 150, there is evidence for a slower decline in the X-rays.




\begin{figure*}
\begin{center}
\includegraphics[clip, angle=-90, width=16cm]{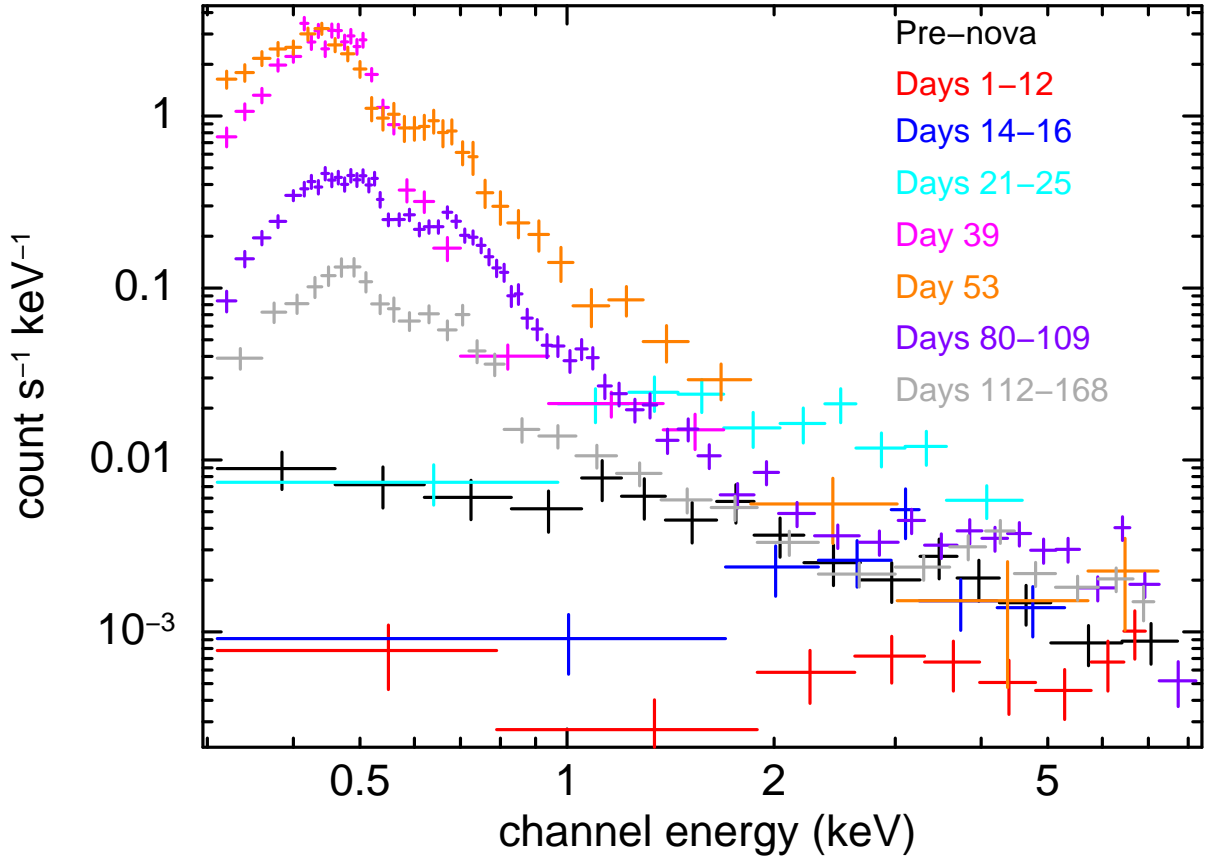}
\caption{A selection of {\it Swift}-XRT spectra of V2491~Cyg taken during the evolution, showing the rise of the hard and then soft spectral components. The spectral resolution of the XRT is $\sim$~180~eV FWHM at 1~keV.}
\label{spec}
\end{center}
\end{figure*}

Figure~\ref{spec} shows a small sample of the XRT spectra obtained throughout the evolution of the nova. Before day 25, the emission was relatively hard, though the source soon evolved into a SSS (significantly softer than the pre-outburst observation shown in Figure~\ref{1bin}). This strong soft emission, which is obvious after about day 36, also varies with time, with some evidence for absorption and emission features visible in Figure~\ref{spec}. Two observations including grating data were also obtained with {\it XMM-Newton} during the SSS phase on days 40 and 50 after outburst (Ness et al. 2008b,c, paper in prep.)

Simple spectral models consisting of an optically thin hard X-ray emission component, together with a blackbody (BB) during the SSS phase, were fitted to the spectra shown in Figure~\ref{spec} and the results are given in Table~\ref{fits}, which also includes a fit to a later spectrum not plotted for clarity. The top panel of Figure~\ref{bands} shows the hard and soft X-ray light-curves on the same scale, highlighting the vast difference between them: almost all the variability is in the soft band, corresponding to the BB emission.  Figure~\ref{bands} also shows how the temperature of the BB component initially increases as the absorbing column decreases, before both become approximately constant around day 57. N$_{\rm H}$ levels off at this time at the best fit value determined from the pre-nova data ($\sim$~2.2~$\times$~10$^{21}$~cm$^{-2}$; Ibarra et al. 2009). It must be noted that BBs should be considered only as an empirical parametrization of the data over the observed (0.3--10~keV) energy band (Krautter et al. 1996); detailed atmosphere models would be required for the modelling to be physically realistic. We therefore also applied a static local thermal equilibrium (LTE) stellar atmosphere model (MacDonald \& Vennes 1991) to our spectra, but found that the fits were significantly worse, caused mainly by incorrect modelling of the oxygen edge (see Figure~\ref{modelcomp}); we also note that the atmosphere model does not fit temperatures above 86~eV, which is insufficient for our data. These fits did, however, reveal similar changes in the parameters as when using the BB approximation (increasing temperature until around day 60; rapidly decreasing luminosity and radius until around this same time); Figure~\ref{atmos} shows the fit results from the CO model, since Figure~\ref{modelcomp} demonstrates that this model fits the data slightly better than the ONe one.

BB fits are known to underestimate the temperature slightly and can overestimate the predicted luminosity by orders of magnitude when applied to low spectral resolution and limited bandwidth (such as {\it ROSAT} PSPC spectra; Krautter et al. 1996). Figure~3 of McDonald \& Vennes (1991) demonstrates that the relationship between the BB and atmosphere temperatures is non-linear, implying that the slopes of the trends found here should be considered
as general indications of the likely true trend. The soft X-ray flux of WD models is very sensitive to the chemical composition of the atmosphere and without high spectral resolution data, it is hard to be confident in the applicability of any given atmosphere model.

Since BBs provide a commonly used approximation which can indicate trends in temperature and absorption, these are the results we concentrate on in this paper. We expect that the underlying trends found from these fits -- if not the absolute luminosity and radius measurements -- are reliable. The radius and luminosity values in Figure~\ref{bands} have therefore been normalised to a peak value of $\sim$~1 by dividing each by a factor of 280.




The spectra after about day 80 are much better fitted by the inclusion of a partial covering component, with a column of between 1--2~$\times$~10$^{23}$~cm$^{-2}$ and a covering fraction of approximately 0.75; without this, the temperature for the optically thin component was unfeasibly high. The inclusion of this component is a spectral fitting expedient which allows the higher energy flux to be well-fitted; we do not infer the creation of a new absorbing screen.
 Where the statistics permit, the later-time spectra show evidence for an iron line, which is accounted for by the optically thin component having a temperature of around 5~keV. The spectra are harder than expected for such a temperature, hence the improvement in the fit by incorporating a partial covering absorber. It seems possible  that we are seeing reflection in the system as accretion resumes (e.g., Done \& Osborne 1997; see Section~\ref{flick}), which could account for both the excess emission at high energies and the iron line.


Figure~\ref{spec} shows evidence for spectral features at low energies (probably oxygen and nitrogen), superimposed on the BB continuum. The XMM high-resolution spectra show a deep {\sc oi} edge, similar to RS~Oph (Ness et al. 2007), and will be explored in more detail by Ness et al. (in prep).

Figure~\ref{bands} demonstrates that the change in hardness ratio seen in Figure~\ref{1bin} is mainly due to the brightening of the soft component, with the emission above 1.5~keV fading only gradually after day 50; the hard and soft bands do not appear to be directly related. Even after day 57 (until $\sim$~day 150) the soft emission is still fading much more rapidly than the hard band. Fitting these decays with a power-law [f~$\propto$~(t-t$_0$)$^{-\alpha}$, taking t$_0$ to be the time of outburst] yields  $\alpha_{\rm soft}$~=~3.08~$\pm$~0.04, compared to  $\alpha_{\rm hard}$~=~0.82~$\pm$~0.14. The data are more variable than a steady power-law decay, however (see Section~\ref{flick}).

\subsubsection{UV data}

Under close examination, the UVOT data show occasional drop-outs, where the source appears to be briefly fainter than expected from adjacent orbits of data. We believe this to be an instrumental effect, since a nearby field star sometimes shows the same behaviour. 

\begin{figure*}
\begin{center}
\includegraphics[clip, width=16cm]{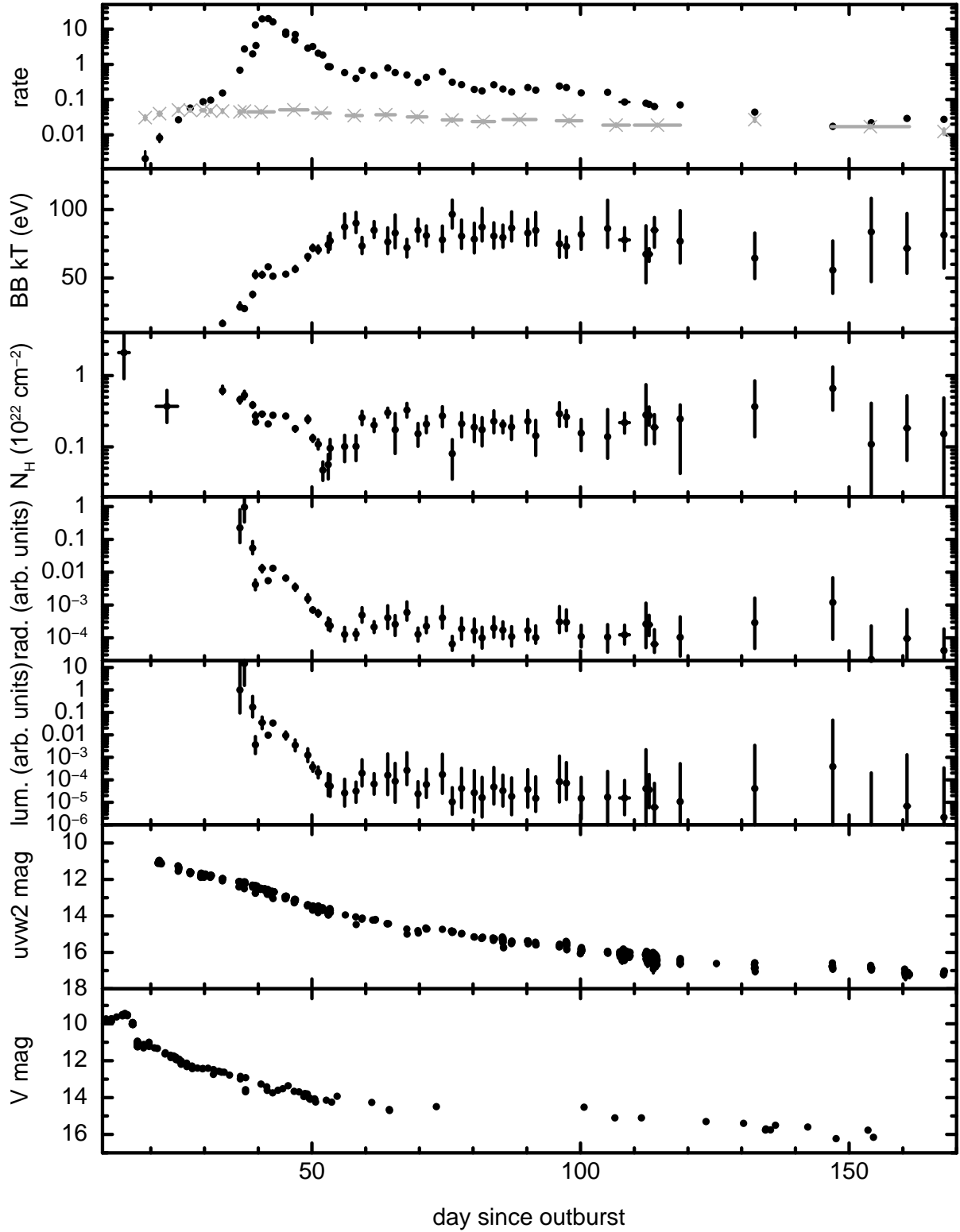}
\caption{Light-curves and derived spectral parameters for V2491~Cyg. The top panel shows the X-ray count-rate in the soft (0.3--1.5~keV; black circles) and hard (1.5--10~keV; grey crosses) bands. The second and third panels show the blackbody temperatures and absorbing column obtained by fitting the spectra. The fourth panel plots the change in radius of the emitting blackbody, while the fifth panel shows the bolometric luminosity estimate from the BB component. Both the radius and luminosity are plotted in normalised units: the values in terms of 10$^9$~cm and the Eddington luminosity have been scaled down by a factor of 280. 
The UVOT uvw2 data are plotted in the sixth panel, with data before day 20 being excluded, since the detector was saturated; the uncertainties are smaller than the marker size. AAVSO $V$-band data are shown in the bottom panel.}
\label{bands}
\end{center}
\end{figure*}

\begin{figure}
\begin{center}
\includegraphics[clip, angle=-90, width=8cm]{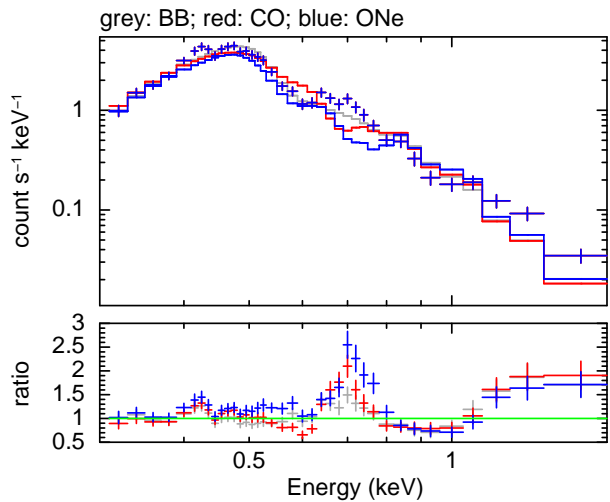}
\caption{A comparison of the BB fit with two atmosphere models, for a CO nova and an ONe nova respectively. The BB model is a much better fit around 0.7~keV. This may arise due to inappropriate LTE or chemical assumptions in the atmosphere models.}
\label{modelcomp}
\end{center}
\end{figure}

\begin{figure}
\begin{center}
\includegraphics[clip, angle=-90, width=8cm]{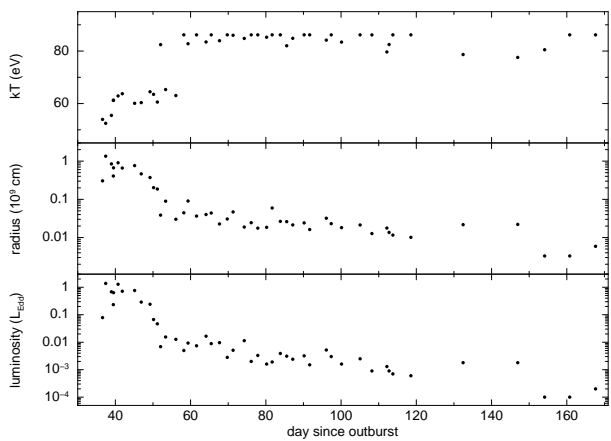}
\caption{Fitting the spectra with a CO nova atmosphere model reveals the same trends as using a BB (Figure~\ref{bands}), but the fits are statistically worse. The model does not extend to high enough temperatures to account for the upper limits on the temperatures, so no error bars have been plotted.}
\label{atmos}
\end{center}
\end{figure}

The decay of the UV brightness steepened at the time of the peak X-ray flux (fitting the UVOT flux, the power-law decay changes from $\alpha$~=~2.02~$\pm$~0.01 to 4.67~$\pm$~0.02 on day 40.41$^{+0.05}_{-0.04}$), before flattening again
around the end of the main SSS outburst ($\alpha$~=~2.73~$\pm$~0.01 after day 57.1~$\pm$~0.2). A flattening around this later time is also clear in the optical data in Figures~\ref{1bin} and \ref{bands}. This is discussed further in Section~\ref{corr}.

\begin{table*}
\begin{center}
\begin{tabular}{lcccc}
\hline
Day & Mekal kT (keV) &  BB kT (eV) & N$_{\rm H}$ (10$^{22}$ cm$^{-2}$) & $\chi^2$/dof\\
\hline
1--12 & 7.8$^{+7.3}_{-2.8}$ & -- & 4.8$^{+2.3}_{-1.7}$  & C-stat = 109/89 \\

14--16 & $>$1.5  & -- &  2.1$^{+72.9}_{-1.2}$ & 4.1/3\\

21--25 & $>$5.1 & -- & 0.37$^{+0.25}_{-0.15}$ & 8.4/7\\
39 & 1.56$^{+0.96}_{-0.52}$ & 38~$\pm$~2 & 0.38$^{+0.04}_{-0.03}$ & 52.7/32\\
53 & 2.53$^{+1.50}_{-0.66}$ & 77$^{+6}_{-4}$ & 0.06$^{+0.02}_{-0.03}$ & 91.4/60\\  


80--109$^{a}$ & 5.4$^{+3.1}_{-2.4}$ & 78$^{+3}_{-2}$ & 0.24~$\pm$~0.01 & 169.4/107\\  

112--168$^{a}$ & 5.4$^{+40}_{-2.3}$ & 75$^{+7}_{-5}$ & 0.22~$\pm$~0.05 & 71.6/54\\

206--236$^{a}$ & 4.9$^{+8.9}_{-1.5}$ & 74$^{+32}_{-19}$ &0.25$^{+0.33}_{-0.19}$  & C-stat = 69/86\\

\hline
\end{tabular}

\caption{Fits to the selection of spectra shown in Figure~\ref{spec}, together with the data after day 206 described in the text. Cash statistics (Cash 1979) were required for the first and last spectra. Errors are given at the 90\% confidence level. $^{a}$ These fits include an additional partial covering component; see Section~\ref{post:x} for more details.}
\label{fits}
\end{center}
\end{table*}

\section{Discussion}
\label{disc}

Prior to the launch of {\it Swift} (Gehrels et al. 2004), only a few novae had been observed in X-rays, and then only to a limited extent (Kuulkers et al. 2006; Krautter 2008). {\it Swift}, however, has allowed us to follow a number of novae throughout much of their outbursts in X-rays (e.g. RS Oph -- Bode et al. 2006; Osborne et al. 2009b; V458 Vul -- Drake et al. 2008a,b,c; CSS 081007:030559+054715 -- Schwarz et al. 2008; Beardmore et al. 2008; Osborne et al. 2009a) and continues to reveal new and fascinating behaviour.

High-cadence {\it Swift} observations of V2491~Cyg from one day after the discovery of the nova have revealed a complex X-ray evolution, with hard optically thin emission shortly after the outburst, a supersoft phase and the source fading back to approximately quiescence. High density photometry also shows that there are breaks in the UV and optical light-curve decay.

The distance to V2491~Cyg is not well known. Ibarra et al (2009) and Takei et al. (2009) both adopt the
value of 10.5~kpc proposed by Helton et al. (2008) on the basis of the
Della Valle and Livio (1995) maximum magnitude versus rate of decline
relationship, assuming E(B-V) = 0.43 (Rudy et al. 2008). Use of the
alternative Downes and Duerbeck (2000) mean absolute magnitude at 15 days,
$<$M$_{v,15}$$>$ = $-$6.05~$\pm$~0.44 for a sample of 28 novae, suggests a distance of
10~$\pm$~2~kpc, although the hump in the lightcurve at around 15 days
reduces confidence further. We note that neither of these relationships is verifed
for recurrent novae (RNe; see later), and that our adopted distance of 10.5~kpc places the nova at z~=~800~pc, beyond the $<$500~pc Galactic scale height of white dwarfs (as
discussed by Benacquista and Holley-Bockelmann 2006).

\subsection{Phases of X-ray emission}
\label{x}

V2491~Cyg shows at least three separate phases of X-ray emission, as has been found for other novae (e.g. Krautter 2008). During the first five weeks after the initial nova explosion, the X-ray emission was hard. Such behaviour has been previously seen in V838~Her (O'Brien, Lloyd \& Bode 1994) and V1974~Cyg (Balman et al. 1998) for example, and is interpreted as emission from shocks caused by components within the ejecta having different velocities. In the case of V2491~Cyg, the faintness of the earliest detections post-nova may be partly due to absorption by the ejected nova shell, as is suggested by the high absorbing column needed by the spectral fits (Table~\ref{fits}). Between about days 14 and 36 after the outburst, the X-ray emission gradually brightens; during this time, as shown in Table~\ref{fits} and Figure~\ref{bands}, the absorbing column decreases. This can be explained by the expansion and thinning of the ejecta which were thrown off the WD surface during the nova explosion. Between about days 30 and 57, N$_{\rm H}$ decreases by a factor of 5--6, implying that the radius of the absorbing material has expanded by up to a factor of 2.5 or so.

The second phase is that of the SSS. Novae are expected to undergo a phase of super-soft emission towards the end of their outburst (e.g., MacDonald, Fujimoto \&
Truran 1985). The predicted nuclear burning at constant luminosity decreases the envelope
mass, possibly with further mass loss due to a stellar wind; the effective
radius of the photosphere is expected to shrink and the corresponding photospheric
temperature to rise. SSS phases have previously been seen in a small number
of novae (e.g. U Sco -- Kahabka et al. 1999; V1494 Aql -- Drake et al. 2003; V382 Vel -- Orio et al. 2002;
V4743 Sgr -- Ness et al. 2003; V1974 Cyg -- Krautter et al. 1996, Balman,
Krautter \& {\" O}gelman 1998; RS~Oph -- Osborne et al. 2009b; Page et al. 2008b; Page et al. in prep; V458~Vul -- Drake et al. 2008a,b,c; CSS~081007:030559+054715 -- Schwarz et al. 2008), however our dataset on V2491~Cyg is outstanding in its complete coverage of this phase.

The X-ray evolution of V2491~Cyg is quite different from some of these
other SSS novae, though. RS~Oph, V458~Vul and
CSS~081007:030559+054715 each showed a rapid variation phase in their
X-ray emission, which is absent from the light-curve of
V2491~Cyg. Although, as Figure~\ref{wiggles} demonstrates, we do see short-timescale variations in this source, the other novae show dramatic count-rate increases and decreases over a matter of days. Here, we see a fast-slow-fast trend in increasing X-ray
emission, from around days 20--27, 27--36 and 36--40 respectively. The
X-ray count-rate peaked around day 40, and remained
constant at $\sim$~20~count~s$^{-1}$ for two days, before starting to
fade. The temperature of the blackbody during this decline is
higher than at the peak of the X-ray emission by about a factor of two
(Figure~\ref{bands}).

The third stage of X-ray emission sees the decline of the SSS, with the hard component again becoming prominent, since it fades much less rapidly than the SSS emission. Underlying hard, optically thin emission was always required to fit the XRT spectra presented here (see Table~\ref{fits}), as well as the BB accounting for the temporary supersoft component; this optically thin component models the 1.5--10~keV data which are shown by the grey crosses in Figure~\ref{bands}, at a count-rate consistent with the pre-nova level.  The X-ray luminosity of this component is $\sim$~3~$\times$~10$^{34}$~erg~s$^{-1}$ during the peak of the outburst, corresponding to a counts-to-luminosity conversion of approximately 7~$\times$~10$^{35}$~erg~count$^{-1}$.

After about day 150, the decline in the X-ray emission has almost ended. At this stage, the optically thin component used to model the high energies contributes about half the count-rate measured below 1.5 keV.


Figure~\ref{1bin} shows the evolution of the count-rate and hardness ratio throughout the {\it Swift} observations. The steady increase in flux, and softening of the emission, after day 14 are due to both the increase in the BB temperature (possibly due to an increasing transparency of the photosphere, allowing hotter regions at smaller radii to become visible) and the decrease in absorbing column (see Figure~\ref{bands}). The following decrease in X-ray count-rate after $\sim$~day 42 appears to be predominantly caused by the apparent shrinking of the emitting area, indicated by the decrease in the best-fit BB radius; energy released by possible settling of the cooling envelope and mass falling back to the WD surface could contribute to the BB temperature remaining constant. It is not clear whether the apparent shrinkage we see is actually due to infall or atmospheric thinning.



As previously mentioned, the absolute value of the luminosity from the BB fits is likely to be an overestimate caused by the lack of line and edge features in the BB model. Similarly, the fact that the minimum radius (after about day 57) derived from the BB model fits is smaller than the white dwarf ($\sim$~5~$\times$~10$^8$~cm) could be a consequence of the BB parametrization. This is, however, also the case for the model atmosphere, as shown in Figure~\ref{atmos}; the very small radius of the WD in the model atmosphere may also be exposing the limitations of this model. Alternatively, the small radius could be hinting towards emission from a restricted region of the WD surface; however, it is difficult to conceive of how this could occur in the framework of nuclear burning. We note that Hachisu \& Kato (2009) invoke B$_{\rm WD}$~=~3~$\times$~10$^{7}$~G on the WD surface and WD-companion star reconnection. It is improbable that such a magnetic field could restrict or suppress nuclear burning (and, hence, soft X-ray emission) to a small region of the WD surface (see, e.g., King, Osborne \& Schenker 2002). In addition, we show in Section~\ref{mag} that the WD in V2491~Cyg is unlikely to be magnetic.

Orio et al. (2001) and Gonz{\' a}lez-Riestra et al. (1998) both conclude that most classical novae decline in bolometric luminosity 1--5~years after outburst, which highlights the fact that V2491~Cyg turned off very rapidly. A similar quick turn-off was found for RS~Oph, a known recurrent nova with a high mass ($>$1.3~M$_{\odot}$) WD. The {\it Swift} data for RS~Oph (e.g., Osborne et al. 2009b; Page et al. 2008b) showed that the SSS phase lasted for around 60 days in total, with the duration of the constant X-ray count-rate stage being $\sim$~15~days. V2491~Cyg may be a recurrent nova and so also  have a high mass WD, a subject we return to in Section~\ref{recurrent}.


Another perspective is provided by considering the phase of rapid radius
decline seen in both the BB and model atmosphere parametrizations, from day $\sim$~40 to $\sim$~57, as a quasi-static optically thick collapse
of the WD atmosphere, with the Kelvin-Helmholtz timescale, t$_{\rm KH}$~=~GM$\Delta$M/2RL, where $M$ is the WD mass and $\Delta M$ is the infalling envelope mass. Using 1$\msun$ as an estimate of the WD mass and 5~$\times$~10$^8$~cm for the radius, this implies an atmosphere mass of the order 10$^{-7} \msun$ if the blackbody parameters from day 40 in Figure~\ref{bands} are taken at face value (or 10$^{-8} \msun$ if the model atmosphere values in Figure~\ref{atmos} are used). This is only a crude approximation but, with this view of the behaviour, the phase of bright soft X-ray emission corresponds not to the constant luminosity nuclear burning phase, but rather to the subsequent relaxation.




\subsection{UV, optical and multi-wavelength analysis}
\label{corr}

Baklanov, Pavlenko \& Berezina (2008) found a 0.0958 day periodicity in their optical ($V$ and $R$ band) observations, although the alias period of 0.10595 day is also possible. No such modulation is seen in the UVOT data, with an upper limit on the amplitude for either period of $\sim$~7~per~cent; neither are they detected in the XRT data to limits of 12~per~cent (days 60--110) and 16~per~cent (days 110--236). This limit of 7~per~cent corresponds to 0.08 magnitudes, making the UVOT insufficiently sensitive to measure the 0.03 -- 0.05 magnitude variations reported by Baklanov et al. (2008). 

The second peak in the optical light-curve of V2491~Cyg is reminiscent of the rebrightening seen in V2362~Cygni (Lynch et al. 2008b) and a small number of other novae (classical and recurrent -- see Bonifacio et al. 2000), though occurred much more rapidly after outburst, after only about 15 days. It is possible that the optical rebrightening in V2362~Cyg occurred at the same time as the increase in X-ray flux, as appears to have also been the case for V2491~Cyg.

There are two points of inflexion in the light-curve at which noticeable changes occur in the decay rate in more than one waveband. About day 40, the X-ray count-rate peaks; at this same time, there is a `blip' in the rise of the BB temperature and a steepening in the decay of the UV flux. Around day 57, the decline in both the X-rays and UV slows (and they begin to fade at a similar rate to each other: $\alpha_{\rm X, soft}$~$\sim$~3.0 and $\alpha_{\rm UV}$~$\sim$~2.7) and the BB temperature and N$_{\rm H}$ level out, becoming close to constant. While the reasons for these correlated changes in the light-curve declines at 40 and 57 days are unknown, it is clear that there is some form of energy linkage between the X-ray and UV bands. 


Novae in outburst are expected to pass through a stage of constant bolometric luminosity which is predicted to have a duration inversely related to the white dwarf mass, although observations indicate the reality is more complicated (Starrfield et al. 1991; Vanlandingham et al. 2001; Gonz{\' a}lez-Riestra, Orio \& Gallagher 1998 and references therein). However, as Figures~\ref{bands} and \ref{atmos} show, the luminosity of V2491~Cyg seems to decline throughout the main outburst period, although days 36--37 at the start of the SSS phase show luminosities which are constant within the error bars. Hence, we see no real observational evidence for a constant bolometric luminosity phase while V2491~Cyg is a SSS. This may be due to the maximum photospheric temperature (i.e., when the supersoft emission moves into the detectable X-ray band) being reached after this phase (e.g., Paczy{\' n}ski \& {\. Z}ytkow 1978; Iben 1982; Driebe et al. 1998).

\subsection{Flickering}
\label{flick}

\begin{figure}
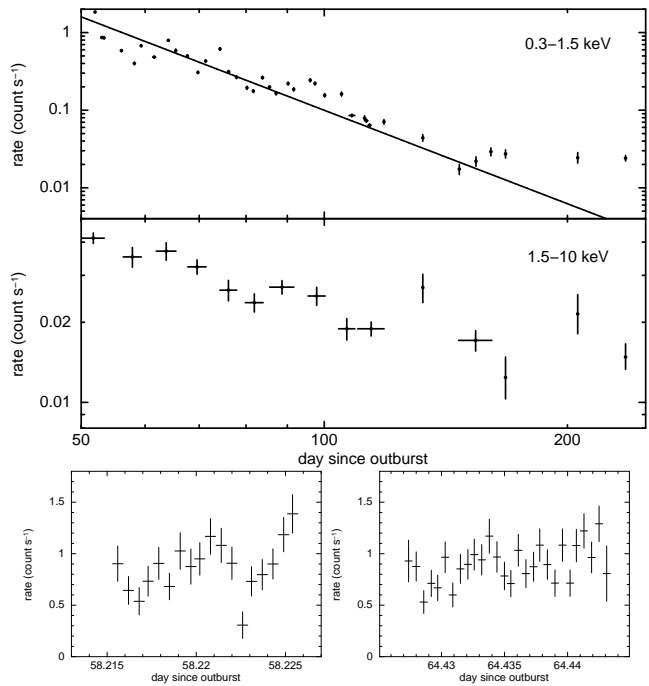

\begin{center}
\includegraphics[clip, angle=-90, width=8.5cm]{flickering_model.ps}
\includegraphics[clip, angle=-90, width=4cm]{038b_flick.ps}
\includegraphics[clip, angle=-90, width=4cm]{041c_flick.ps}
\caption{The top panel shows the soft and hard X-ray count-rate evolution of V2491~Cyg after day 57. Substantial flickering  over periods of days is evident in both bands, though is much stronger at lower energies. The two frames in the lower panel show sample 0.3--10~keV light-curves with 50~s bins, demonstrating flickering over a shorter timescale as well.}
\label{flickering}
\end{center}
\end{figure}

Beyond about day 57, after the main SSS outburst had ended, significant flickering in the X-ray emission became apparent, both on timescales of a few days and five to ten minutes. Figure~\ref{flickering} plots close-ups of the fluctuating X-ray emission. As the figure shows, the light-curves appear to have power-law-like decays until at least day 150, so we use this model as a purely empirical means of characterising the intensity decline. However, in both the 0.3--1.5 and 1.5--10~keV bands, a steady decay is statistically unacceptable, with $\chi^2_{\nu}$~$\sim$~46 for the soft band, and $\chi^2_{\nu}$~$\sim$~2.5 for the harder data. A power-law fit to the UV data over this same time period also leads to a poor $\chi^2_{\nu}$ of $\sim$~46. Once the general fading trend has been subtracted from the data, there is no correlation between the variations in the  soft and hard bands (the Spearman Rank test gives a correlation probability of 51~per~cent); the data were compared using identical timebins (those plotted for the hard band in  Figure~\ref{flickering}).
The standard deviation of the trend-removed soft points is more than an order of magnitude greater than that of the hard points. 


Short-term flickering is often thought to be a signature of accretion and so can be used to place a time limit for the disc to be rebuilt. The viscous timescale in the accretion disc is given by 

\begin{equation}
\label{eq:visc}
t_{\rm v} \sim  \frac{R^2}{\alpha c_{\rm s} H}
\end{equation} 

where $\alpha$ is the viscosity ($\sim$~0.1), c$_{\rm s}$ -- the sound speed ($\sim$~10$^{6}$~cm~s$^{-1}$), R -- the radius and H -- the scale height (H/R~$\sim$~0.05), taking typical values from, e.g., Frank, King \& Raine (1985). From the data presented here, the flickering becomes noticeable around day 57. If this is the time needed for the disc to reach back down to the WD surface, it suggests that the accretion disc was destroyed out to a radius of $\sim$~2~$\times$~10$^{10}$~cm. 




From Kepler's law, the binary separation is $\sim$~7~$\times$~10$^{10}$~(M$_{\rm tot}$/1.5$\msun$)$^{1/3}$~cm, where M$_{\rm tot}$ is the system mass and the orbital period is taken to be the photometric period of 0.0958~day measured by Baklanov et al. (2008). As an accretion disc can be expected to have a radius of about half the binary separation, substantial, but not complete, destruction of the disc is suggested, consistent with the re-establishment of accretion on the observed timescale.

\subsection{A recurrent nova?}
\label{recurrent}

Classical novae (CNe) are, by definition, only observed to go into outburst once, although they are expected eventually to repeat on timescales of 10$^{3-5}$ years. Thus, an object may be reclassified from a classical to recurrent nova (that is, one which repeats over a matter of decades) if multiple eruptions are observed. 

V2491 Cyg is a fast nova, shown both by its expansion velocities (up to $\sim$~4860~km~s$^{-1}$) and the rapid fading of the optical emission (see Figures~\ref{1bin} \& \ref{bands}; Tomov et al. 2008a). Fast novae are thought to contain massive white dwarfs, with the U~Sco type  of RNe being among the fastest observed (Warner 2008; Sekiguchi 1995).

Although CNe remain X-ray emitters for some time after their explosions, previously only one other has been identified as an X-ray source pre-outburst: 
V2487 Oph, another very fast nova, is coincident with 1RXS~J173200.0-1934 (Hernanz \& Sala 2002), detected by the {\it ROSAT} all-sky survey. Novae must, of course, accrete prior to explosion, but their typically low accretion rates mean that any X-ray emission would be weak and difficult to detect at typical nova distances. 
V2487~Oph was suggested to be a strong candidate for being a recurrent nova (Hachisu et al. 2002) because of both the rapid decline in the optical and the presence of a plateau phase during the decline (which are features common for the U~Sco subclass of RNe). More recently, Pagnotta, Schaefer \& Xiao (2008) discovered a previous outburst of the system in 1900, confirming its recurrent nature.

There are certain characteristics which the members of the class of RNe share, although they are, in general, quite an inhomogeneous group. It is thought that the WD mass may be larger for RNe than CNe and that the accretion rate is also higher (Starrfield 1989). 
The secondary stars in these binary systems are thought to be evolved (Warner 1995; Anupama \& Miko{\l}ajewska 1999; Orio et al. 2005), in comparison to the Main Sequence stars in CNe systems.
RNe with giant secondaries tend to be fast nova systems, with broad optical emission lines which narrow over time, as well as high excitation lines (Anupama \& Miko{\l}ajewska 1999).

Tomov et al. (2008b) proposed that V2491~Cyg may also be a recurrent nova, citing the presence of similar spectral characteristics (e.g., classification as a `He/N' nova, the line profiles and the lack of forbidden lines) to two recurrent novae, U Sco and V394 CrA, as well the existence of pre-nova X-ray emission (first reported by Ibarra et al. 2008). A high inter-outburst accretion rate, suggested by the relatively high X-ray luminosity between outbursts (allowing the detection of the source in X-rays), would be expected to be a signature of RNe, since they need to replenish the ejected material on a much shorter timescale than CNe.

\begin{figure}
\begin{center}
\includegraphics[clip, width=9cm]{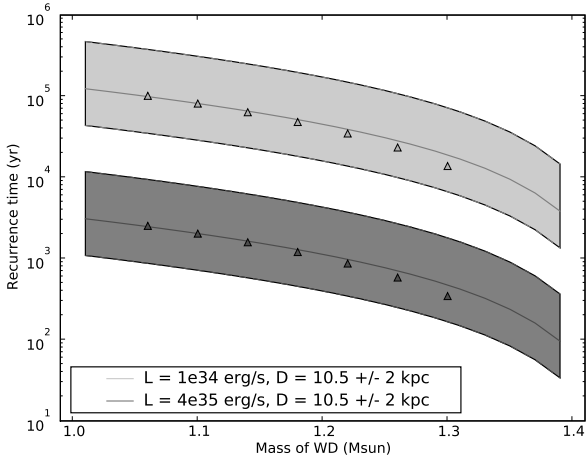}
\caption{Nova recurrence time as a function of WD mass for V2491~Cyg. The solid curves represent calculations using the WD mass-radius relation of Nauenberg (1972), while the triangles represent calculations based on the models of Althaus et al. (2005; we assume a temperature of 5~$\times$~10$^{4}$~K in this case, although we note that the results are not strongly temperature dependent). The dashed lines indicate the $10.5\pm2$ pc error envelope in the distance, as well as the uncertainty in the accretion efficiency. 
The upper set of curves assume an inter-outburst luminosity of $10^{34}$ erg s$^{-1}$ while the lower set 
of curves assume $4\times 10^{35}$ erg s$^{-1}$ (see Ibarra et al. 2009).}
\label{trec}
\end{center}
\end{figure}

The inter-outburst luminosity estimates ($L_X \sim 10^{34}$ -- $4 \times 10^{35}$
erg s$^{-1}$) reported by Ibarra et al. (2009) allow us to make estimates 
of the recurrence time for V2491~Cyg. The critical mass for nova ignition
on a WD is (Starrfield 1989)
\begin{equation}
\label{eq:mcrit}
M_{\rm crit} = \frac{4\pi R_{\rm WD}^4 P_{\rm crit}}{G M _{\rm WD}},
\end{equation} 
where $P_{\rm crit} \sim 10^{20}$ dyne cm$^{-2}$ is the critical ignition 
pressure. Relating the inter-outburst luminosity to the mean accretion
rate of the WD between outbursts ($\dot{M}_{\rm WD}$)
\begin{equation}
\label{eq:lacc}
L_{\rm acc} = \frac{\epsilon GM_{\rm WD} \dot{M}_{\rm WD}}{R_{\rm WD}}
\end{equation}
(where $\epsilon$ is an efficiency parameter, between $\sim$~0.1 and 0.5, related to the spin of the WD and other uncertainties in the global energy budget; Popham \& Narayan 1995)
and assuming a WD mass-radius relation allows us to estimate the 
nova recurrence time $t_{\rm rec} \sim M_{\rm crit}/\dot{M}_{\rm WD}$. Figure~\ref{trec} shows $t_{\rm rec}$ as a function of WD mass for the mass-radius 
relationship of Nauenberg (1972)
\begin{equation}
\label{eq:lacc}
\frac{R_{\rm WD}}{R_\odot} = 1.12\times 10^{-2} \biggl [ \biggl ( \frac{M_{\rm WD}}{1.44 M_\odot} \biggr )^{-2/3} - \biggl ( \frac{M_{\rm WD}}{1.44 M_\odot} \biggr )^{2/3} \biggr ]^{1/2}
\end{equation}
(solid curves) and the recent calculations of Althaus et al. (2005) for massive white dwarfs (triangular points). 
It is clear from the figure that the WD in V2491~Cyg must be massive ($\go$~1.3$\msun$) if it is to be 
recurrent on any reasonable timescale. If the WD were to have a mass of $\sim$~1.4$\msun$, then V2491~Cyg could recur within a matter of decades, similar to RS~Oph. However, since no previous nova outburst has actually been detected in this system, we suggest that, if V2491~Cyg is indeed a recurrent nova, it is
is more likely to be similar to V2487~Oph, recurrent over a timescale of $\sim 100$ years. However, at present we only have hints that the mass of the WD is high and it is possible that the accretion rate measured by Ibarra et al. (2009) is not typical of the inter-outburst interval; thus, a short recurrence time is still speculative.


\subsection{A magnetic white dwarf?}
\label{mag}

Ibarra et al. (2009) consider that the pre-nova X-ray spectra of V2491~Cyg may be similar to those of 
magnetic cataclysmic variables. Building on this, Hachisu \& Kato (2009) develop a model
in which V2491~Cyg is a polar system with a strongly magnetic WD ($B_{\rm WD} \sim$~3~$\times$~ 10$^7$ G)
to explain the secondary maximum in the optical lightcurve. A polar contains a strongly magnetic
WD whose spin is closely locked to the binary orbital period ($P_{\rm orb}$). This 
period-locking is thought to come about because the interaction between the magnetic fields of the two stars
is able to overcome the spin-up torque of the accreting matter. The synchronisation condition can be 
expressed as (e.g.\ King, Frank \& Whitehurst 1991)
\begin{equation}
\label{eq:synch}
\frac{\mu_1 \mu_2}{a^3} > \frac{2\pi \dot{M}_{\rm WD} b^2}{P_{\rm orb}}
\end{equation}
where $\mu_1 \sim B_{\rm WD} R_{\rm WD}^3$ and $\mu_2 \sim B_2 R_2^3$ are the magnetic moments of 
the WD and the secondary star respectively
and $b$ is the distance of the WD from the inner Lagrange point. Taking $P_{\rm orb} \sim 0.0958$ day 
(Baklanov et al. 2008), $M_{\rm WD} \sim 1.3 M_\odot$, $M_2 \sim 0.25 M_\odot$, $\dot{M}_{\rm WD} \sim
10^{-7}$ $M_\odot$ yr$^{-1}$ (Ibarra et al. 2009) and $\mu_2 \sim 10^{35}$ G cm$^3$ and using 
\begin{equation}
\label{eq:ba}
\frac{b}{a} = 0.500 - 0.227 \log{\frac{M_2}{M_{\rm WD}}},
\end{equation}
where $a$ is the binary separation (Warner 1976), we find a critical field strength $B_{\rm WD, crit}
\sim$~500--1000 MG (dependent mainly on the radius of the WD) for synchronisation. The WD in V2491~Cyg would need a surface magnetic field
strength of at least this value to be a phase-locked polar system and this would place it among 
the most magnetic WDs known in binaries. We note that this value is an 
order of magnitude larger than that adopted by Hachisu \& Kato (2009) in their model for the 
secondary maximum in the optical light curve. 

One of the fundamental observable properties of magnetic cataclysmic variable stars is periodic, pulsed
X-ray emission on the WD spin period ($P_{\rm spin}$). This pulsed X-ray emission comes about because the 
the stream of mass from the secondary star is magnetically channelled towards the 
magnetic poles of the WD. On accretion, the gas is shocked, producing strong X-ray emission.
The X-rays are then strongly modulated on $P_{\rm spin}$
because of self-occultation and absorption effects caused by the rotation of the WD. 
The condition for magnetic channelling is that the magnetic pressure should be at least of the same
order as the ram pressure of the accreting gas
\begin{equation}
\frac{B^2}{8\pi} \sim \rho v^2
\end{equation}
where $\rho$ is the gas density and $v$ its velocity. Using the equation of continuity for the 
magnetically channelled region $\dot{M}_{\rm WD} \sim  \pi R_{\rm WD}^2 \rho v$ and 
$v \sim (GM_{\rm WD}/ R_{\rm WD})^{1/2}$, we estimate a critical magnetic field strength for channelling
as $\lo$~1~MG. A WD with a surface magnetic field of this order or greater
would be expected to show strongly modulated X-ray emission during the inter-outburst accretion 
phases. Clearly, if V2491~Cyg satisfies the polar synchronisation condition, the accretion flow must 
be channelled and pulsed X-ray emission should be observable. We repeat that the X-ray and UV emission were not detectably modulated at the 0.0958~day period found by Baklanov et al. (2008) in their optical data; this detected modulation shows that the system cannot be pole-on, an orientation which could otherwise be used to explain the lack of rotational modulation in the X-rays. The above considerations suggest that a magnetic WD is unlikely in this system. 


\section{Summary}

The evolution of V2491~Cyg (Nova Cyg 2008 No. 2) has been closely followed by {\it Swift} in the X-ray and UV bands from day 1--236 after the detection of the nova in April 2008.
The X-ray emission was seen to rise, pass through a super-soft state and then to decline, although with additional changes in rate of decay. By the time of the last observation, the X-ray count-rate had returned to approximately the pre-outburst level.

In the absence of suitable atmosphere models, we characterise the soft X-ray evolution by time-resolved blackbody spectral fits, showing variations in the radius and temperature. Although a cursory inspection would imply that the X-ray and UV light-curves are independent of one another, our work shows that the decline in the UV emission alters at the times of changes in the behaviour of the soft X-ray emission. In additional to the large scale changes in the X-ray light-curve, variability is seen over a few hundred second timescales. Late-time flickering may be caused by the resumption of accretion.

The high X-ray flux of the pre-nova suggests that V2491~Cyg could be another member of the small class of recurrent novae and thus have a high ($\go$~1.3 $\msun$) WD mass.


\section{ACKNOWLEDGMENTS}
\label{ack}

We acknowledge with thanks the variable star observations from the AAVSO International Database contributed by observers worldwide and used in this research.
We thank Stephen Holland, Martin Still and Mat Page for advice regarding the UVOT data analysis and the Swift PI, Neil Gehrels, together with the Swift science and mission operations teams for their support of these observations. We also thank Jeremy Drake for providing a prompt and useful referee report.

KLP, PAE, JPO, APB and RLCS acknowledge STFC support.

\end{document}